\documentclass[aps,prd,reprint,twocolumn,superscriptaddress,showpacs]{revtex4-1}
\usepackage{amsfonts}
\usepackage{mathrsfs}
\usepackage{amsmath}% needed for subequations
\usepackage{color}
\usepackage{natbib}
\usepackage{textcomp}
\usepackage{graphicx}
\usepackage{bm}% bold maths
\usepackage{amssymb}
\usepackage{xspace}
\usepackage{epstopdf}
\usepackage{dcolumn}% Align table columns on decimal point
\usepackage{longtable}
\usepackage{multirow}
\usepackage[colorlinks=true, letterpaper=true, pdfstartview=FitV, linkcolor=blue, citecolor=blue, urlcolor=blue]{hyperref}
\usepackage{float}
\usepackage{bm}

\makeatletter

\newcommand{\Rmnum}[1]{\expandafter\@slowromancap\romannumeral #1@}
\makeatother
\begin{document}

\title{Antiferromagnetic and Electric Polarized States in Two-Dimensional Janus Semiconductor Fe$_2$Cl$_3$I$_3$}

\author{Zhen Zhang}
\email{These authors contributed equally to this work.}
\affiliation{School of Physical Sciences, University of Chinese Academy of Sciences, Beijng 100049, China}

\author{Jing-Yang You}
\email{These authors contributed equally to this work.}
\affiliation{School of Physical Sciences, University of Chinese Academy of Sciences, Beijng 100049, China}

 \author{Bo Gu}
 \email{gubo@ucas.ac.cn}
 \affiliation{Kavli Institute for Theoretical Sciences, and CAS Center for Excellence in Topological Quantum Computation, University of Chinese Academy of Sciences, Beijng 100190, China}
\affiliation{Physical Science Laboratory, Huairou National Comprehensive Science Center, Beijing 101400, China}

\author{Gang Su}
\email{gsu@ucas.ac.cn}
\affiliation{School of Physical Sciences, University of Chinese Academy of Sciences, Beijng 100049, China}
\affiliation{Kavli Institute for Theoretical Sciences, and CAS Center for Excellence in Topological Quantum Computation, University of Chinese Academy of Sciences, Beijng 100190, China}
\affiliation{Physical Science Laboratory, Huairou National Comprehensive Science Center, Beijing 101400, China}

\begin{abstract}
Two-dimensional (2D) Janus semiconductors with mirror asymmetry can introduce novel properties, such as large spin-orbit coupling (SOC) and normal piezoelectric polarization, which have attracted a great interest for their potential applications. Inspired by the recently fabricated 2D ferromagnetic (FM) semiconductor CrI$_3$, a stable 2D (in x-y plane) antiferromagnetic (AFM) Janus semiconductor Fe$_2$Cl$_3$I$_3$ with normal sublattice magnetization ($\bm{m}$$\parallel$$\bm{z}$) is obtained by density functional theory calculations. By applying a tensile strain, the four magnetic states sequentially occur: AFM with $\bm{m}$$\parallel$$\bm{z}$ of sublattice, AFM with $\bm{m}$$\parallel$$\bm{xy}$ of sublattice, FM with $\bm{m}$$\parallel$$\bm{xy}$, and FM with $\bm{m}$$\parallel$$\bm{z}$. Such novel magnetic phase diagram driven by strain can be well understood by the spin-spin interactions including the third nearest-neighbor hoppings with the single-ion anisotropy, in which the SOC of I atoms is found to play an essential role. In addition, the electric polarization of Fe$_2$Cl$_3$I$_3$ preserves with strain due to the broken inversion symmetry. Our results predict the rare Janus material Fe$_2$Cl$_3$I$_3$ as an example of 2D semiconductors with both spin and charge polarizations, and reveal the highly sensitive strain-controlled magnetic states and magnetization direction, which highlights the 2D magnetic Janus semiconductor as a new platform to design spintronic materials.
\end{abstract}
\maketitle

%%%%%%% Main text %%%%%%%%%%%%%%%%%%%%%

\section{\uppercase\expandafter{\romannumeral1}.    Introduction}
%{\color{blue}{\em Introduction}}---
Two-dimensional (2D) materials, such as graphene, transition metal dichalcogenides and black phosphorus~\cite{Novoselov2004,Yu2007,Novoselov2005,Jin2009,Mak2010,Li2014}, have attracted tremendous attention due to their excellent electrical, optical and acoustic properties. Although many efforts have been devoted to investigating 2D materials, the 2D semiconductors with intrinsic magnetism are still rare~\cite{Bonilla2018,Wang2016,Long2017,Zhou2016,Zheng2018,Ersan2019}. Recently, the successful synthesis of intrinsic ferromagnetic (FM) semiconductor monolayer CrI$_{3}$~\cite{Huang2017} and bilayer  CrGeTe$_{3}$~\cite{Gong2017} with the Curie temperature of 45K and 28K, respectively, has attracted much attention on 2D magnetic semiconductors. However, potential applications of these magnets in spintronics, high Curie temperature above room temperature is highly required. In addition, the large magnetic anisotropy is needed to stabilize the magnetism in 2D systems, according to Mermin-Wagner theorem~\cite{Mermin1966}. The large magnetic anisotropy is predicted in the technetium based 2D magnetic semiconductors~\cite{You2020}. A useful approach to tune magnetism and Curie temperature is by strain~\cite{Webster2018,Huang2018,Zheng2019,Dong2019,Baskurt2020,Ersan2019a,Iyikanat2018,Sarikurt2018,Vatansever2019,Shen2019,Wu2019}, which can be realized by bending flexible substrates, elongating an elastic substrate, exploiting the thermal expansion mismatch and so on~\cite{Conley2013,He2013,Wang2015,Hui2013,Plechinger2015,Castellanos-Gomez2013,Roldan2015}.The band gap can also be modified by strain. For example, a transition from the direct band gap semiconductor to a metal was proposed to occur in MoS$_2$ monolayer with a tensile strain up to 15$\%$~\cite{Scalise2011}. The topological properties, such as the Weyl half-semimetal~\cite{You2019} and the room-temperature quantum anomalous Hall effect~\cite{You2019b} are recently proposed in the 2D ferromagnetic semiconductors. Other than FM materials, antiferromagnetic (AFM) spintronics began to take a shape, because AFM materials can not only be used as an assistant material, such as pinning layers to control the magnetization direction of the adjacent ferromagnetic layers, but also can work as a memory~\cite{Marti2014,Wadley2016,Olejnik2018}. Furthermore, the spin seebeck effect in antiferromagnets MnF$_{2}$ has recently been observed in the experiment~\cite{Wu2016}. Therefore, the investigation of the FM and AFM spintronics becomes necessary and interesting.

Among various 2D materials, the 2D Janus materials are very attractive. Compared to their protypes, Janus materials have broken symmetries, and thus can induce many intriguing properties, such as large spin-orbit coupling (SOC), piezoelectricity, polarization, etc.~\cite{Yang2019,Dong2017,Guo2017b,Cheng2013,Yin2018,Xu2020,Kandemir2018}. The first graphene-based Janus material was graphone, where the Dirac cone was opened with a small gap and the FM was obtained~\cite{Zhou2009}. Substituting one sulfur layer with selenium in GaS, the piezoelectric coefficient in Ga$_{2}$SSe was enhanced as large as four times~\cite{Guo2017a}. Recently, some magnetic Janus materials such as VSSe\cite{Zhang2019}, Cr$_2$I$_3$X$_3$\cite{Zhang2019a,Zhang2020}, and V$_2$Cl$_3$I$_3$\cite{Ren2020} were theoretically studied, and they exhibited interesting properties such as large piezoelectricity and valley polarization, enhanced Cuire temperatures.
The 2D Janus material MoSSe~\cite{Lu2017,Zhang2017a}, which has been successfully synthesized recently, not only has a better hydrogen evolution reaction efficiency, but also possesses the topological and ferroelastic properties~\cite{Ma2018} compared with its protype MoS$_{2}$ monolayer~\cite{Er2018}.

In this work, by studying 2D Janus materials M$_2$Cl$_3$I$_3$ (M=3$d$ transition metals) on the basis of the crystal of CrI$_3$, we propose a stable 2D magnetic Janus semiconductor Fe$_{2}$Cl$_3$I$_3$. By means of first-principle calculations, Fe$_{2}$Cl$_3$I$_3$ was found to be a 2D AFM semiconductor with out-of-plane magnetization ($\bm{m}$$\parallel$$\bm{z}$) of sublattice. Due to the charge redistribution caused by different electronegativity of Cl and I atoms and the broken inversion symmetry, Fe$_{2}$Cl$_3$I$_3$ monolayer possesses electrical polarization of about 0.18 e{\AA} and piezoelectricity of about 4.48 pm/V. By applying biaxial tensile strain up to about 15$\%$ on Fe$_{2}$Cl$_3$I$_3$ monolayer, a novel phase diagram with four magnetic states is found: AFM with out-of-plane magnetization of sublattice, AFM with in-plane magnetization of sublattice, FM with in-plane magnetization, and FM with out-of-plane magnetization. The magnetic phase can be well understood by the spin-spin interactions with single-ion anisotropy term, the latter is mainly determined by the spin-orbit coupling of I atoms. Our results demonstrate a strain-controlled magnetic phases of 2D Janus magnetic semiconductors controlled by strain, and thus suggest a promising way to design functional materials.

\section{\uppercase\expandafter{\romannumeral2}. Computational Methods}
Our first-principles calculations were carried out with the Vienna ab initio simulation package (VASP) based on the density functional theory (DFT)~\cite{Kresse1993,Kresse1996}. The interactions between nuclei and electrons were described by the projector augmented wave (PAW) method~\cite{Bloechl1994}, and the generalized gradient approximation (GGA) in the form proposed by Perdew, Burke, and Ernzerhof (PBE) \cite{Perdew1996} was used to describe the electron exchange-correlation functional. In order to prevent the unphysical interlayer interactions, we build a 20{\AA} vacuum. The cutoff energy was set to be 520 eV, and the K-meshes for structure optimization and self-consistent calculations is $9\times9\times1$ and $15\times15\times1$ $\Gamma$-centered Monkhorst-Pack grid~\cite{Monkhorst1976}, respectively. The structure optimization of atomic positions and the lattice vectors were done until the maximum force on each atom was less than 0.0001 eV/${\AA}$, and the total energy was converged to 10$^{-8}$ eV. During the optimization, the conjugate gradient (CG) scheme were employed. To account for the correlation effects of Fe 3$\textit{d}$ electrons, the GGA + SOC + $\textit{U}$ ($\textit{U}$ = 4 eV which is typical for 3$\textit{d}$ electrons) was
used in most of our calculations, and the effect of different $\textit{U}$ values was also investigated. The phonon frequencies were obtained by the density functional perturbation theory (DFPT) as implemented in the PHONOPY code \cite{Togo2015} using a $2\times2\times1$ supercell. And the molecular dynamics (MD) simulations in the canonical (NVT) ensemble were performed in a $3\times3\times1$ supercell at 300K with a Nos\'{e} thermostat.

%whose mirror symmetry was broken with two kinds of halogens atoms Cl and I stay different sides of the Fe atoms. the electronic and magnetic properties were also explored
%{\color{blue}{\em DFT results}}
\section{\uppercase\expandafter{\romannumeral3}.    Structural, magnetic and electronic properties}
\subsection{A. Crystal structures and stability}
\begin{figure*}[!hbt]
  \centering
  \includegraphics[scale=0.15,angle=0]{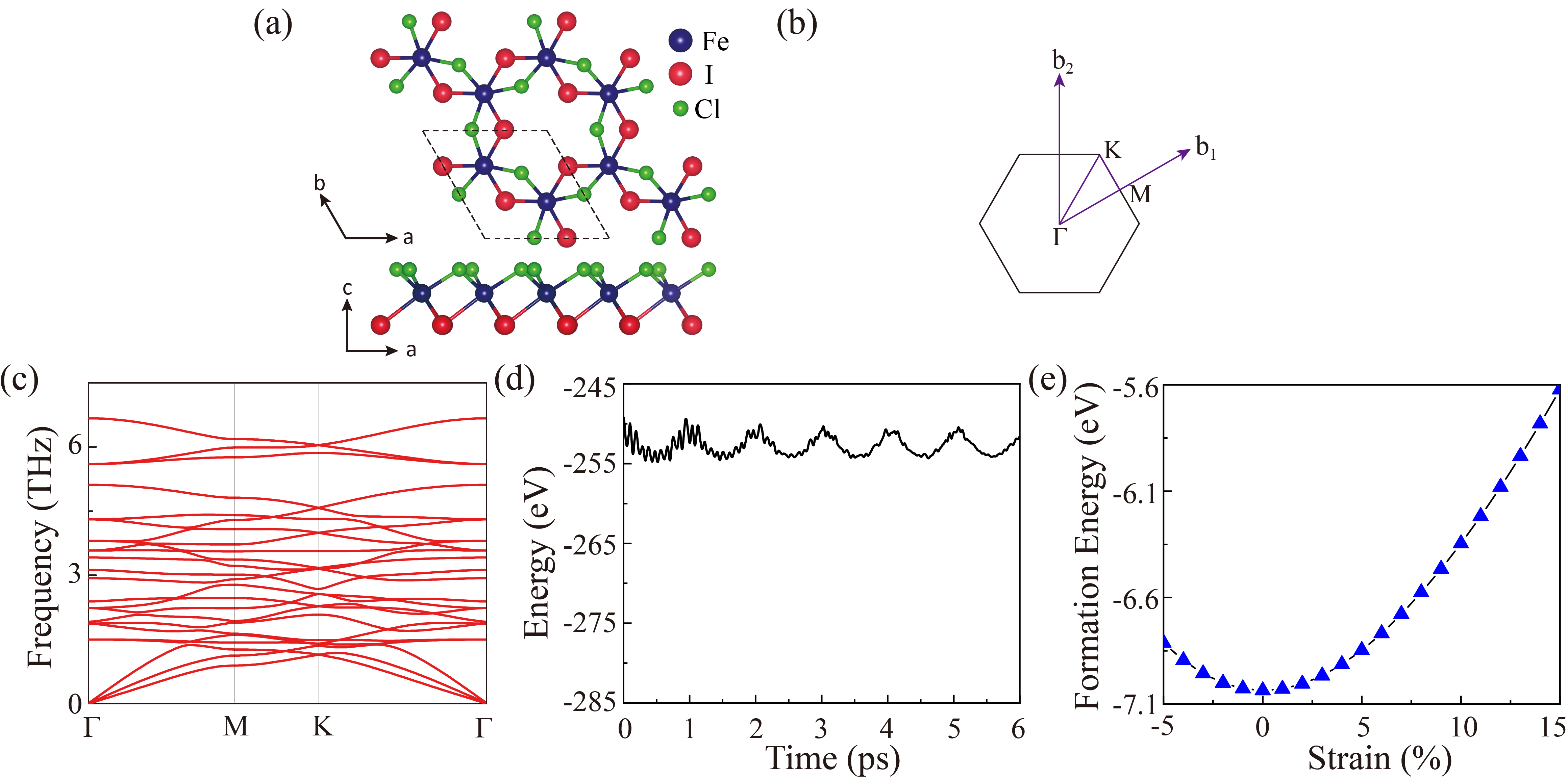}\\
  \caption{(a) Top and side views of the 2D Janus material Fe$_{2}$Cl$_3$I$_3$. (b) The first Brillouin zone with high symmetry points labeled. (c) Calculated phonon spectra. (d) MD simulations of Fe$_2$Cl$_3$I$_3$ at 300K for 6ps with a time step of 3fs. (e) Formation energy as a function of applied strain.}\label{fig1}
\end{figure*}

The crystal structure of 2D Janus Fe$_{2}$Cl$_3$I$_3$ is shown in Fig.\ref{fig1}a, where the Fe atoms are sandwiched by two different halogen atomic layers Cl and I. 2D Fe$_{2}$Cl$_3$I$_3$ with the broken inversion symmetry belongs to the \emph{P}31m (No.157) space group. Each primitive cell contains one formula units, and the Fe atoms locate in the center of the distorted octahedron consisting of three Cl atoms and three I atoms, and form a honeycomb lattice.

To examine the stability of 2D Fe$_{2}$Cl$_3$I$_3$, its formation energy was calculated. The formation energy is defined as $E_f = E_{Fe_2Cl_3I_3} - 2E_{Fe} - 3/2E_{Cl_2} - 3/2E_{I_2}$, where $E_{Fe_2Cl_3I_3}$ is the energy of the Fe$_2$Cl$_3$I$_3$ monolayer, $E_{Fe}$ is the energy of bulk bcc Fe, $E_{Cl_2}$ and $E_{I_2}$ are the energies of Cl$_2$ and I$_2$ molecular dimers, respectively.
As shown in Table \ref{tab:formation}, the formation energy of Fe$_2$I$_3$Cl$_3$ is between FeCl$_3$ and FeI$_3$. The negative value of $E_f$=-7.04 eV per primitive cell indicates an exothermic reaction. Inspired by the successful synthesis of MoSSe in experiments ~\cite{Lu2017,Zhang2017a}, we propose a similar synthetic scheme to fabricate Janus Fe$_3$Cl$_3$I$_3$ as follows

~~~~~4FeI$_3$ + 3Cl$_2$ $\rightarrow$ 2Fe$_2$Cl$_3$I$_3$ + 3I$_2$,

\noindent
where the energy difference between products (Fe$_2$Cl$_3$I$_3$, I$_2$) and reactants (FeI$_3$, Cl$_2$) was -5.33 eV, and the negative value suggests the feasibility of this synthetic scheme.

The phonon spectra of Fe$_{2}$Cl$_3$I$_3$ monolayer were calculated as shown in Fig.\ref{fig1}c, where no imaginary frequency mode in the whole Brillouin zone indicates that Fe$_{2}$Cl$_3$I$_3$ monolayer is dynamically stable. Moreover, after 6ps MD simulation with a time step of 3fs as shown in Fig.\ref{fig1}d, no structural changes occur, and Fe$_2$Cl$_3$I$_3$ still keeps a honeycomb lattice, revealing the thermal stability of Fe$_2$Cl$_3$I$_3$. The optimized lattice constant is 6.717{\AA}. To investigate the mechanical property of monolayer Fe$_2$Cl$_3$I$_3$, the Young's modulus was calculated to be 11.4 N/m, which is much smaller than that of MoS$_2$ (180 N/m)~\cite{Bertolazzi2011} and graphene (342 N/m)~\cite{Politano2015,Lee2008}, and MoS$_2$ and graphene could suffer from 11\% and 13\% strain, respectively. The lower Young's modulus indicates the possible applications of Fe$_2$Cl$_3$I$_3$ monolayer under a larger tensile strain~\cite{Bertolazzi2011}. Moreover, the formation energy of Fe$_2$Cl$_3$I$_3$ as a function of applied strain, which is defined as $(a-a_0)/a_0$ is shown in Fig.\ref{fig1}e. One may see that it changes continuously from compressed strain (-5$\%$) to tensile strain (15$\%$) and keeps negative values, revealing the stability of Fe$_2$Cl$_3$I$_3$ under the applied strain.

\begin{table}[t]
\begin{center}
\caption{The formation energy \textit{E$_f$}  (in eV) per primitive cell for Fe$_2$Cl$_3$I$_3$, FeCl$_3$ and FeI$_3$ monolayers calculated by GGA + SOC + \textit{U} (\textit{U} = 4 eV) method. Per primitive cell for FeCl$_3$ and FeI$_3$ is Fe$_2$Cl$_6$ and Fe$_2$I$_6$.}\label{tab:formation}
\centering
\setlength{\tabcolsep}{7mm}{
\begin{tabular}{cccccc}
\hline
         &  Fe$_2$Cl$_3$I$_3$ &  FeCl$_3$  & FeI$_3$  \\
\hline
 E$_f$   &      -7.04         & -8.89      & -4.37    \\
\hline
%\bottomrule
\end{tabular}}
\end{center}
\end{table}
\subsection{B. Antiferromagnetic ground state}
The magnetic ground state of Fe$_{2}$Cl$_3$I$_3$ was studied by comparing the total energy of different spin configurations: FM, N\'{e}el AFM, stripy AFM, zigzag AFM and paramagnetic (PM) configurations. Table \ref{tab:magnet} lists the total energy per Fe$_{2}$Cl$_3$I$_3$ unit cell relative to the ground states. In contrary to FeCl$_3$ and FeI$_3$ monolayers, which possess FM ground state as shown in Table \ref{tab:bare}, Fe$_2$Cl$_3$I$_3$ has the ground state of zigzag AFM with out-of-plane magnetization. The energy difference between the ground state and the zigzag AFM with in-plane magnetization of sublattice is about 3.7 meV.

\begin{table}[t]
\begin{center}
\caption{The total energy per unit cell for Fe$_2$Cl$_3$I$_3$ monolayer (in meV, relative to the total energy of zigzag AFM along $\boldsymbol{z}$-axis magnetization) for several spin configurations of Fe atoms calculated by GGA + SOC + \textit{U} (\textit{U} = 4 eV) method.}\label{tab:magnet}
\begin{tabular}{cccccccc}
\hline
Zigzag AFM($\bm{m}$$\parallel$$\bm{z}$)~~Zigzag AFM($\bm{m}$$\parallel$$\bm{x}$)~~Zigzag AFM($\bm{m}$$\parallel$$\bm{y}$) \\
0.0~~~~~~~~~~~~~~~~~~~~~~3.7~~~~~~~~~~~~~~~~~~~~~7.5 \\
\hline
\\
\hline
N\'{e}el AFM($\bm{m}$$\parallel$$\bm{z}$)~~~~Stripy AFM($\bm{m}$$\parallel$$\bm{z}$)~~~~
FM($\bm{m}$$\parallel$$\bm{z}$)~~~~PM\\
~~~~~~~~~~20.1~~~~~~~~~~~~~~~~~~~~~~19.3~~~~~~~~~~~~~~~~~~~~68.9~~~~~~~472.1\\
        \hline
	\end{tabular}
\end{center}
\end{table}

\begin{figure*}[!hbt]
  \centering
  \includegraphics[scale=0.18,angle=0]{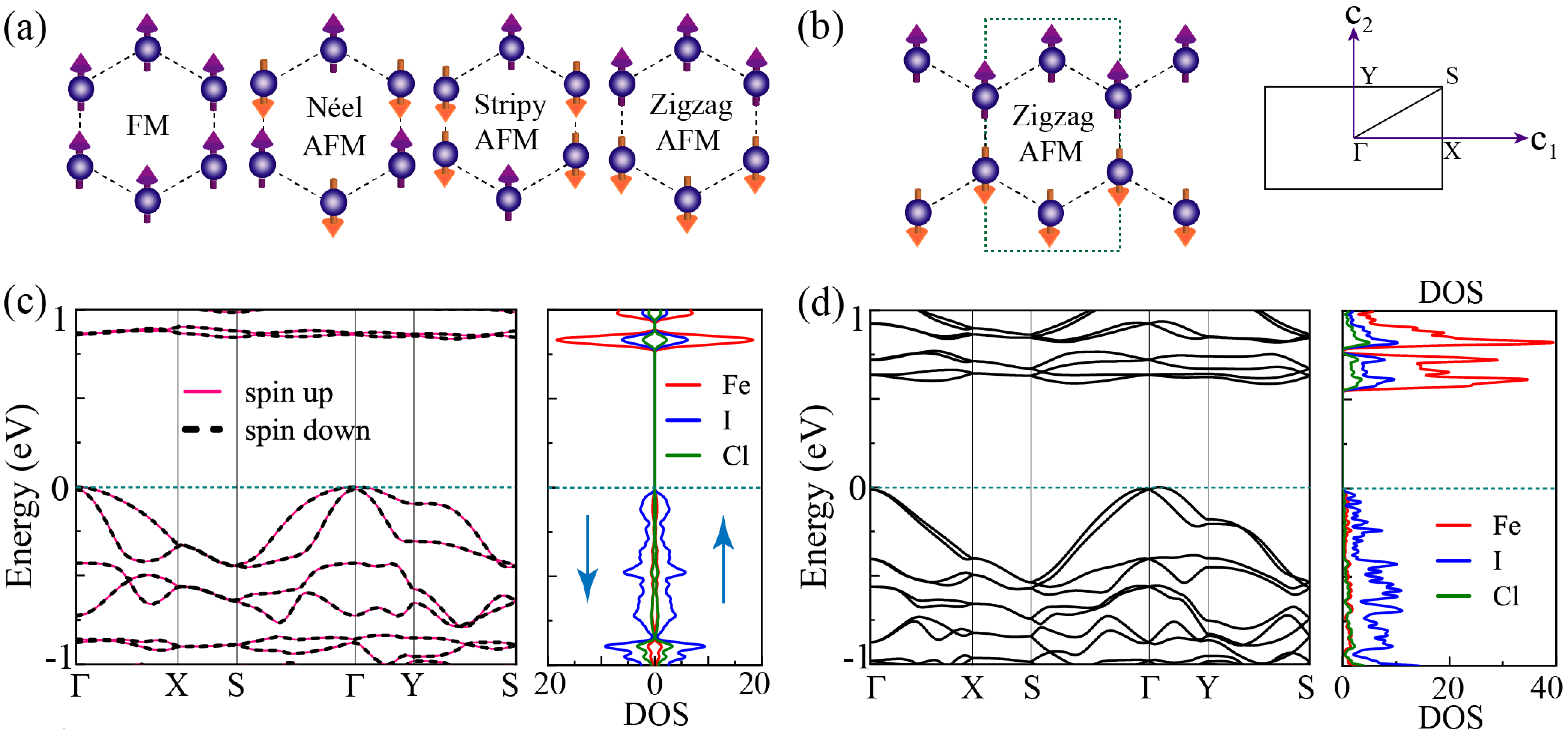}\\
  \caption{(a) Possible spin configurations for Fe atoms: FM, N$\acute{e}$el AFM, Stripy AFM and Zigzag AFM. (b) The unit cell for zigzag AFM spin configuration and its corresponding first Brillouin zone with high symmetry points marked. Band structure and atom-projected DOS of zigzag AFM spin configuration for Fe$_{2}$Cl$_3$I$_3$ monolayer are calculated by (c) GGA + \textit{U} and (d) GGA + SOC + \textit{U} (\textit{U} = 4 eV).}\label{fig2}
\end{figure*}

\subsection{C. Band structure and Electric polarization}

The band structures as well as atom-projected density of states (DOS) of Fe$_{2}$Cl$_3$I$_3$ calculated by GGA + $\textit{U}$ and GGA + SOC + $\textit{U}$ ($\textit{U}$ = 4 eV) are shown in Fig.\ref{fig2}c and Fig.\ref{fig2}d, respectively. There is a large difference between the band structures with and without SOC, where a large SOC effect can be expected in Janus materials due to the breaking of inversion symmetry. Without SOC, Fe$_2$Cl$_3$I$_3$ has an indirect band gap of about 0.83 eV. With including SOC, the band gap decreases to about 0.58 eV. The atom-projected DOS shows that the valence and conduction bands near Fermi level are mainly contributed by I and Fe atoms, respectively. Due to the different electronegativity of Cl and I atoms, the charge redistributes. According to the Bader charge analysis, one I atom gains 0.29 e from the Fe atom, and one Cl atom gains 0.55 e from the Fe atom. So a spontaneous electric polarization along the direction perpendicular to the plane with magnitude of 0.18 e{\AA} was obtained. Thus, the 2D Janus material Fe$_2$Cl$_3$I$_3$ is a rare example of the 2D semiconductors with both spin and charge polarizations.

\subsection{D. Effect of strain}
\begin{figure*}[!hbt]
  \centering
  \includegraphics[scale=0.35,angle=0]{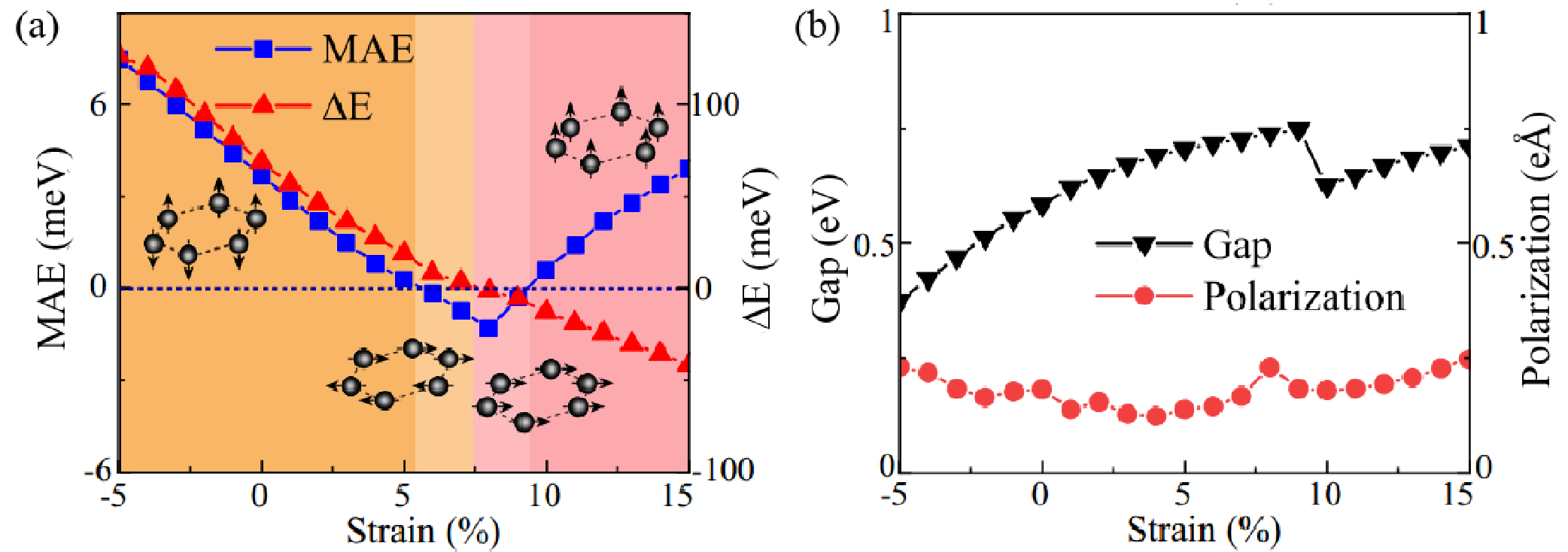}\\
  \caption{Strain-dependent (a) magnetic anisotropy energy (MAE) , energy difference ($\Delta$E) between ferromagnetic and antiferromagnetic configurations and (b) band gap and electric polarization. A positive value of MAE indicates the out-of-plane magnetization, and a positive value of $\Delta$E prefers the antiferromagnetic ground state, otherwise the opposite. The results are calculated by GGA + SOC + \textit{U} (\textit{U} = 4 eV) method. }\label{fig3}
\end{figure*}

\begin{figure*}[!hbt]
  \centering
  \includegraphics[scale=0.60,angle=0]{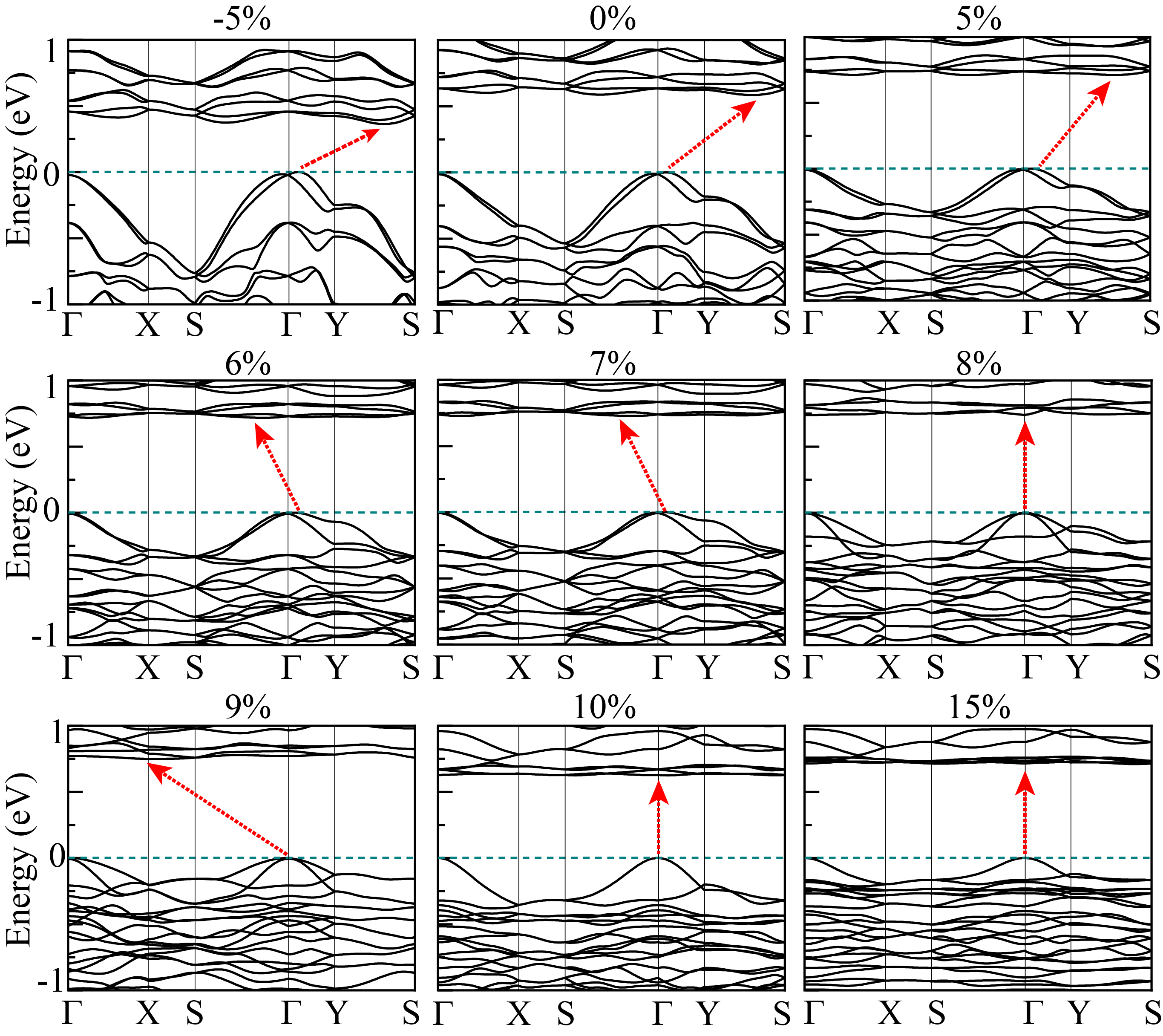}\\
  \caption{Band structures of Janus Fe$_2$Cl$_3$I$_3$ under different strains. The results are calculated  by GGA + SOC + \textit{U} (\textit{U} = 4 eV) method.}\label{fig4}
\end{figure*}

The effects of biaxial strain from compress 5$\%$ to tensile 15$\%$ on the properties of 2D Janus material Fe$_{2}$Cl$_3$I$_3$ are explored. The magnetic anisotropy energy (MAE) defined as the energy difference between the states with in-plane and out-of-plane spin configurations, $\Delta$E defined as energy difference between the FM and AFM, and the band gap and electric polarization as a function of the strain were plotted in Fig.\ref{fig3}. From Fig.\ref{fig3}a, it is noted that, $\Delta$E decreases as the increase of tensile strain, and a phase transition from zigzag AFM to FM occurs with the tensile strain of 7$\%$. Meanwhile, one can observe that MAE changes from positive to negative, and then returns to the positive value, corresponding to the change of magnetization direction from out-of-plane to in-plane, and then back to out-of-plane. The magnetic ground state changed with the strain can be briefly summarized as four steps: zigzag AFM with out-of-plane magnetization of sublattice, zigzag AFM with in-plane magnetization of sublattice, FM with in-plane magnetization, and FM with out-of-plane magnetization. In addition, the magnetic ground states, MAE and band gaps for Fe$_2$Cl$_3$I$_3$, FeCl$_3$ and FeI$_3$ are shown in Table \ref{tab:bare}. It can be observed that the formation energy and band gap of Fe$_2$Cl$_3$I$_3$ lie between those of FeCl$_3$ and FeI$_3$, while the magnetic states of these three materials are quite different. Fe$_2$Cl$_3$I$_3$ has a zigzag AFM ground state with out-of-plane sublattice magnetization, FeCl$_3$ has a FM ground state with out-of-plane magnetization, and FeI$_3$ has a FM ground state with in-plane magnetization. On the other hand, the band gap and electric polarization as a function of strain for Fe$_2$Cl$_3$I$_3$ are calculated by GGA + SOC + \textit{U} (\textit{U} = 4 eV) as shown in Fig.\ref{fig3}b. The band gap and electric polarization preserve with the applied strain. It is interesting to note that as the strain can induce the magnetic phase transition from zigzag AFM to FM phase, the positions of the top of valence band and the bottom of conduction band change under different strains as shown in Fig.\ref{fig4}. The strain can change the overlap and hybridization of atomic orbitals, which could lead to change of the electronic band structures. Fe$_2$Cl$_3$I$_3$ in our paper shows the strain-induced magnetic phase transition between antiferromagnetic ground state and ferromagnetic ground state, and the strain-induced change of electronic band structure between indirect and direct band gap. These strain-induced novel magnetic and electronic properties have not been reported in previous studies of the magnetic Janus materials VSSe\cite{Zhang2019}, Cr$_2$I$_3$X$_3$\cite{Zhang2019a,Zhang2020}, V$_2$Cl$_3$I$_3$\cite{Ren2020}. Moreover, the change of band gap in Fig.\ref{fig3}b shows the discontinuity at tensile strain of 10$\%$, where a magnetic phase transition occurs from in-plane to out-of-plane ferromagnetization.

To achieve the strain effect on two-dimensional Fe$_2$Cl$_3$I$_3$, we could place Fe$_2$Cl$_3$I$_3$ on a two-dimensional substrate, such as MoS$_2$, h-BN and so on. Once a Fe$_2$Cl$_3$I$_3$ primitive cell matches with a $2\times2\times1$ MoS$_2$ cell, whose lattice constant is about 6.36{\AA}, a compressive strain about 5\% is applied. Fe$_2$Cl$_3$I$_3$ will retain the zigzag antiferromagnetic ground state, and the magnetic anisotropy and N\'{e}el temperature will be enhanced. When we match a Fe$_2$Cl$_3$I$_3$ primitive cell with a $3\times3\times1$ h-BN cell, there will be a tensile strain about 12\%, and Fe$_2$Cl$_3$I$_3$ will change to a ferromagnetic ground state.

\begin{table}[t]
\begin{center}
\caption{ Magnetic ground state, magnetic anisotropy energy (MAE) (meV) defined as the energy difference between the in-plane and out-of-plane magnetization configurations, and band gaps (eV) for Fe$_2$Cl$_3$I$_3$, FeCl$_3$ and FeI$_3$, respectively. All above results are calculated by GGA + SOC + \textit{U} (\textit{U} = 4 eV ).}\label{tab:bare}
%\centering
%\begin{tabular}{p{1.5cm}<{\centering}p{1.8cm}<{\centering}p{1cm}<{\centering}p{1cm}<{\centering}}
\setlength{\tabcolsep}{3.5mm}{
\begin{tabular}{cccc}
\hline
           & Magnetic ground state & MAE  & Gap  \\
%&  ~~~~~~          &~~~~a$_0$~~~&~~~GS~~~&~~~~MAE~~~&~~~~Gap~~~~&  \\
\hline
Fe$_2$Cl$_3$I$_3$  & Zigzag AFM ($\bm{m}$$\parallel$$\bm{z}$)  &  3.7   &  0.58    \\
%\hline
FeCl$_3$           &    FM ($\bm{m}$$\parallel$$\bm{z}$)    &  0.2   &  1.87    \\
%\hline
FeI$_3$            &    FM ($\bm{m}$$\parallel$$\bm{x}$)   & -2.4   &  0.49    \\
\hline
	\end{tabular}}
%\centering
\end{center}
\end{table}

\section{Theoretical Analysis}

\begin{figure}[!hbt]
  \centering
  \includegraphics[scale=0.5,angle=0]{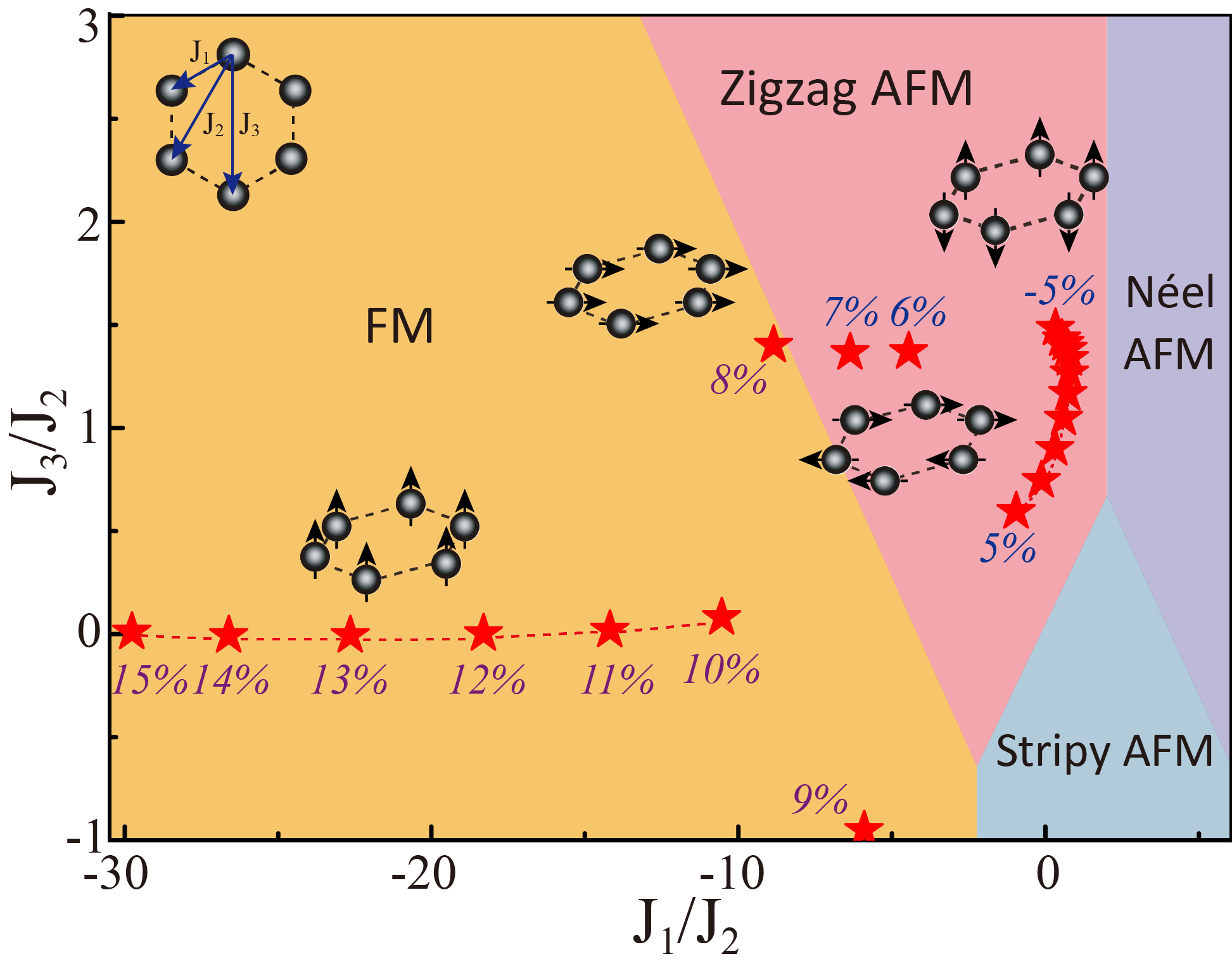}\\
  \caption{Magnetic phase diagram with respect to J$_1$/J$_2$ and J$_3$/J$_2$ for Fe$_2$Cl$_3$I$_3$. The DFT results with different strains are marked with stars along with the plots of the corresponding magnetic ground states.}\label{fig5}
\end{figure}

To better understand the strain-controlled magnetic phase transition in Fe$_2$Cl$_3$I$_3$, we employed a Hamiltonian which can be written as
\begin{equation}
H = H_0 + H_1 ,\label{Hamiltonian}
\end{equation}
\begin{align}
H_0
&     = \sum_{\langle i,j\rangle}J_{1}\boldsymbol{S_i}\cdot\boldsymbol{S_j}+\sum_{\langle\langle i,j\rangle\rangle}J_{2}\boldsymbol{S_i}\cdot\boldsymbol{S_j} + \sum_{\langle\langle\langle i,j\rangle\rangle\rangle}J_{3}\boldsymbol{S_i}\cdot\boldsymbol{S_j},\label{2}
\end{align}
\begin{align}\label{SIA}
H_1
&     = \sum_iA_{xx}S_{i,x}^2+A_{yy}S_{i,y}^2+A_{zz}S_{i,z}^2,
\end{align}
\noindent
where \textit{H$_0$} includes the spin-spin interaction and \textit{H$_1$} contains the single-ion anisotropy. $\boldsymbol{S_i}$ is the spin operator at the $\textit{i}$-th lattice site. J$_1$, J$_2$ and J$_3$ are the nearest-neighbor, next nearest-neighbor and third nearest-neighbor exchange integrals, respectively. $A_{xx/yy/zz}$ represents the
amplitude of single-ion anisotropy along $x/y/z$ direction, and $S_{i,x/y/z}$ is the $x/y/z$ component of the $\textit{i}$-th spin operators. Considering the spin-spin interactions are much larger than single-ion anisotropic energies, for simplicity we ignore the single-ion anisotropy when we estimate J$_1$, J$_2$, and J$_3$ parameters based on DFT results. Although more precise values of exchange parameters in magnetic metals can be obtained using the spin-wave stiffness method~\cite{Liechtenstein1987}, considering the complicated spin-wave behaviors of H$_0$ with three exchange parameters J$_1$, J$_2$, and J$_3$ for the magnetic semiconductor Fe$_2$Cl$_3$I$_3$, it is not readily to estimate these parameters by means of this method. For simplicity, we opt to use the simple method of energy mapping to estimate the exchange parameters J$_1$, J$_2$ and J$_3$,
as many previous works did~\cite{Sarikurt2018,Vatansever2019,Shen2019,Sivadas2015,Xiang2013,Xu2018}. This method can also be adopted to interpret the strain-induced phase transition from AFM to FM in the 2D magnetic semiconductor CrSiTe$_3$~\cite{Sivadas2015}. In order to obtain J$_1$, J$_2$ and J$_3$,
the energies corresponding to four different magnetic configurations: FM, N$\acute{e}$el AFM, stripy AFM and zigzag AFM were expressed as
\begin{equation}
%\left\{
\begin{aligned}
&E_{FM}=(3J_1+6J_2+3J_3)|\boldsymbol{S}|^2+E_0, \\
&E_{N\acute{e}el}=(-3J_1+6J_2-3J_3)|\boldsymbol{S}|^2+E_0,\\
&E_{stripy}=(-J_1-2J_2+3J_3)|\boldsymbol{S}|^2+E_0,\\
&E_{zigzag}=(J_1-2J_2-3J_3)|\boldsymbol{S}|^2+E_0,
\end{aligned}
%\right.
\end{equation}
\noindent
where \textit{E}$_0$ is the energy that is independent of spin configurations. Thus, the magnetic phase diagram with respect to J$_1$/J$_2$ and J$_3$/J$_2$ can be obtained by comparing the energies determined by Eq.(4) for a given set of exchange parameters, as shown in Fig.~\ref{fig5}. On the other hand, from the DFT results of Fe$_2$Cl$_3$I$_3$ with different strains,  the parameters J$_1$, J$_2$, and J$_3$ can be estimated as marked with stars in Fig.\ref{fig5}, where the digital numbers of each star denote the corresponding strain.

As shown in Fig.\ref{fig5}, Fe$_2$Cl$_3$I$_3$ locate at the zigzag AFM and FM phases.
In the FM phase, the magnitude of exchange parameter J$_1$ is much larger than that of J$_2$ and J$_3$, and the model Hamiltonian in Eq.(2) is simplified to the spin-spin interactions with nearest-neighbor coupling J$_1$. While in the zigzag AFM phase, both J$_1$, J$_2$ and J$_3$ are important. Although the jump of digital numbers in the phase diagram correspond to four different magnetic ground states, the phase diagram is obtained from the spin-spin interactions, and can not explain the change of the magnetization directions.

\begin{figure*}[htbp]
  \centering
  \includegraphics[scale=0.53,angle=0]{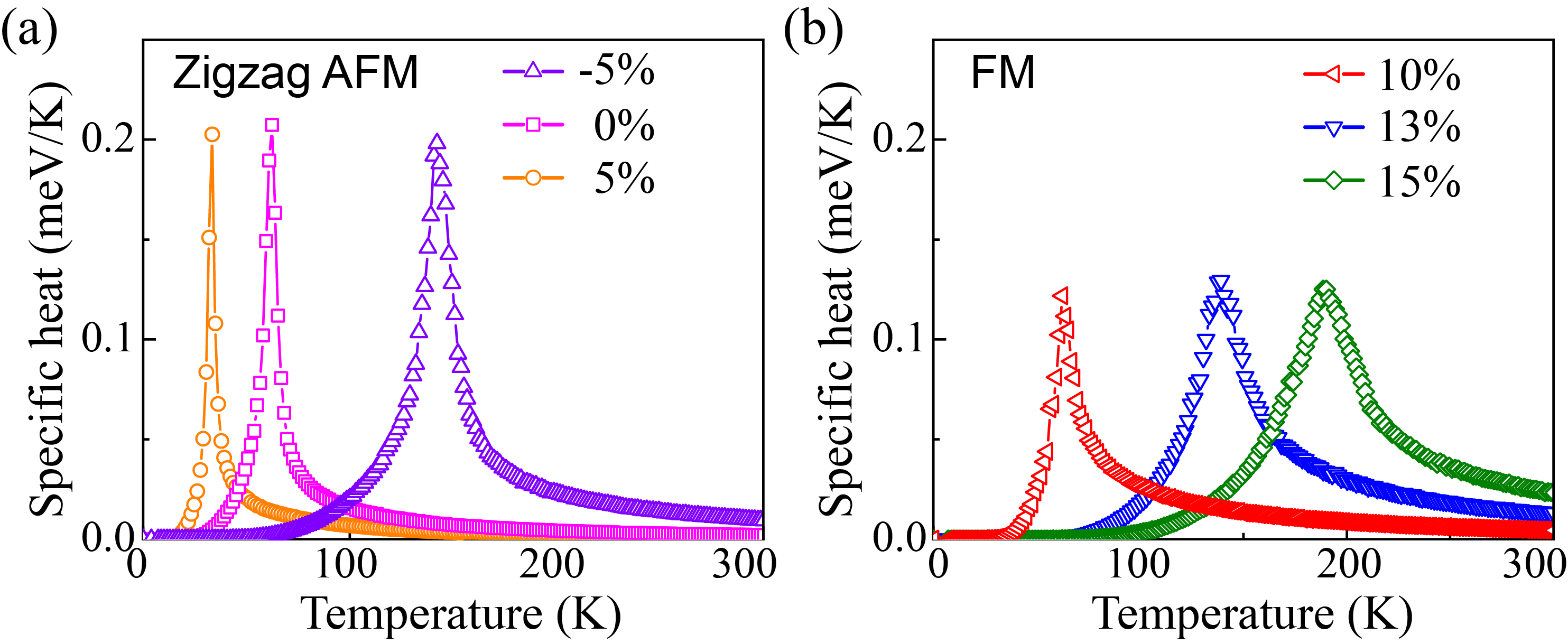}\\
  \caption{ Specific heat as a function of temperature of Janus Fe$_2$Cl$_3$I$_3$ for (a) the zigzag AFM ground state with -5\%, 0, and 5\% strain, and (b) the FM ground sate with 10\%, 13\%, and 15\% strain. The results are calculated by Monte Carlo simulations based on the Ising model.}\label{fig6}
\end{figure*}

Based on the above magnetic exchange parameters J$_1$, J$_2$ and J$_3$, and
keeping the Ising-type interactions in Eq.(2) for simplicity, we use Monte Carlo simulations to estimate the Curie temperature of Janus Fe$_2$Cl$_3$I$_3$ with different biaxial strain. Monte Carlo simulations are performed in a 60$\times$60$\times$1 2D honeycomb lattice with 10$^6$ steps for each temperature calculations. As shown in Fig.\ref{fig6}, for the zigzag AFM ground states, the N$\acute{e}$el temperature decreases from 142K to 33K as the strain changes from -5\% to 5\%. And when the strain changes from 10\% to 15\%, the Curie temperature for the corresponding FM ground states with out-of-plane magnetization increases from 65K to 193K. As expected, the critical temperature decreases as the magnetic ground state approaches to the magnetic phase transition boundary driven by strain.

In order to understand the change of magnetization direction, i.e. the sign of MAE in Fig.~\ref{fig3}a, we consider the single-ion anisotropy (SIA) term as shown in Eq.(3).
According to the second-order perturbation theory, the MAE caused by SIA can be described by \cite{Wang1993,Yang2017}
\begin{align}
E_{SIA}
&     = \lambda^{2}\sum_{o,u}\frac{|\langle\Psi_{u}|L_{x}|\Psi_{o}\rangle|^{2}-|\langle\Psi_{u}|L_{z}|\Psi_{o}\rangle|^{2}}{\varepsilon_{o}-\varepsilon_{u}},
\end{align}
where $\lambda$ is the SOC constant, $L_{z/x}$ represents the angular momentum operator, $\Psi_{u}$ and $\Psi_{o}$ are the wavefunctions of unoccupied and occupied states, respectively, and $\varepsilon_{u}$ and $\varepsilon_{o}$ are the corresponding energy levels. A positive value of $E_{SIA}$ indicates the out-of-plane magnetization, and negative value indicates the in-plane magnetization.

\begin{figure*}[htbp]
  \centering
  \includegraphics[scale=0.18,angle=0]{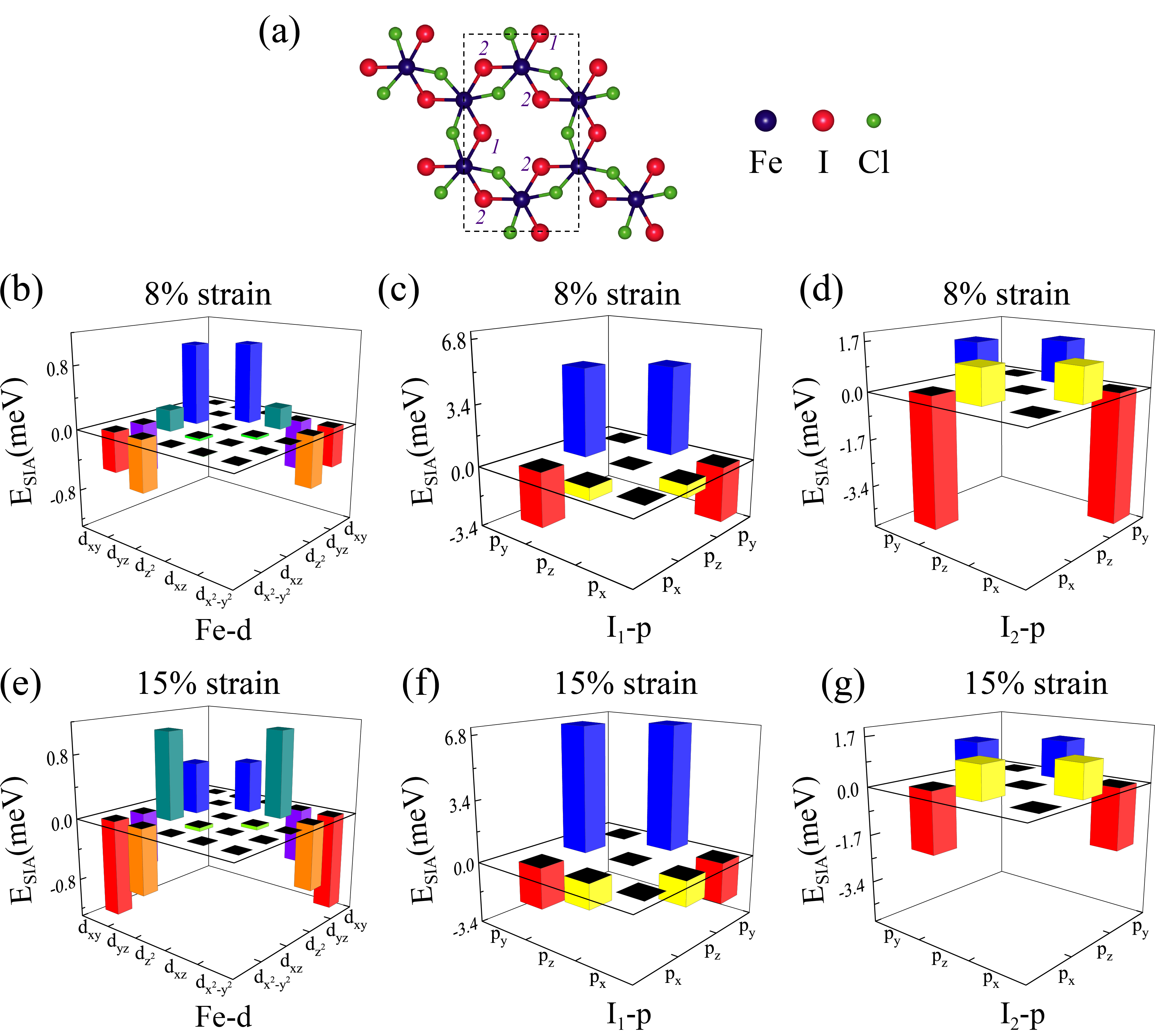}\\
  \caption{(a) The structure of 2D Janus materials Fe$_2$Cl$_3$I$_3$ labeled with two kinds of I atoms. Orbital-resolved single-ion anisotropy energy of Fe and two kinds of I atoms in 2D Janus Fe$_2$Cl$_3$I$_3$ monolayer with (b) (c) (d) 8$\%$ and (e) (f) (g) $15\%$ strain. Different colors represent the contributions to $E_{SIA}$ from different d and p orbitals of Fe and I atoms.}\label{fig7}
\end{figure*}

To unveil the mechanism that leads to the change of $E_{SIA}$ with different strains, we have calculated the orbital-resolved $E_{SIA}$. As an example, the orbital-resolved $E_{SIA}$ of Fe$_2$Cl$_3$I$_3$ with 8$\%$ and 15$\%$ tensile strains, which correspond to FM with magnetization  along the x-axis and FM with magnetization along the z-axis ground states, respectively, was calculated as shown in Fig.\ref{fig7}. By DFT calculations,the main contribution to $E_{SIA}$ is from Fe and I atoms, and I atoms can be classified to two kinds as labeled in Fig.\ref{fig7}a according to their different surroundings when the magnetization is along the x-axis. In the case of 8$\%$ tensile strain, as shown in Table \ref{tab:SIA}, the contributions to $E_{SIA}$ from Fe and two kinds of I atoms are -1.2, 2.5 and -4.1 meV, respectively, giving rise to a total negative $E_{SIA}$ of -2.8 meV. It is consistent with the in-plane magnetization. For the case of 15$\%$ tensile strain, a positive value $E_{SIA}$ of 5.3 meV is obtained, which indicates the out-of-plane magnetization. From Fig.\ref{fig7}, one may observe that $E_{SIA}$ mainly originates from ($p_y$,$p_x$), ($p_y$,$p_z$) and ($p_z$,$p_x$) matrix element of I atoms. As the tensile strain increases from 8$\%$ to 15$\%$, the sign change of $E_{SIA}$ from I atoms gives rise to the sign change of the total $E_{SIA}$, and thus induces the changes of the magnetization direction.

\begin{table}[t]
\begin{center}
\caption{The orbital-resolved single-ion anisotropy energy of Fe and two kinds of I atoms (in meV) in 2D Janus Fe$_2$Cl$_3$I$_3$ monolayer with 8$\%$ and $15\%$ strain, respectively. The results are calculated by GGA + SOC + \textit{U} (\textit{U} = 4 eV) method.}\label{tab:SIA}
\setlength{\tabcolsep}{1.6mm}{
\begin{tabular}{ccccc}
		\hline
		                & ~~~~~Fe-d~~~~~ & ~~~~~I$_{1}$-p~~~~~ & ~~~~~I$_{2}$-p~~~~~ & ~~Total~~\\
		\hline
		      ~~8$\%$~~   &  -1.2   &2.5     &-4.1  & -2.8 \\
             ~~15$\%$~~   &  -2.1   &6.8     &0.6   &  4.1\\
        \hline
	\end{tabular}}
\end{center}
\end{table}

\section{Effect of electronic correlation}

\begin{figure*}[htbp]
  \centering
  \includegraphics[scale=0.45,angle=0]{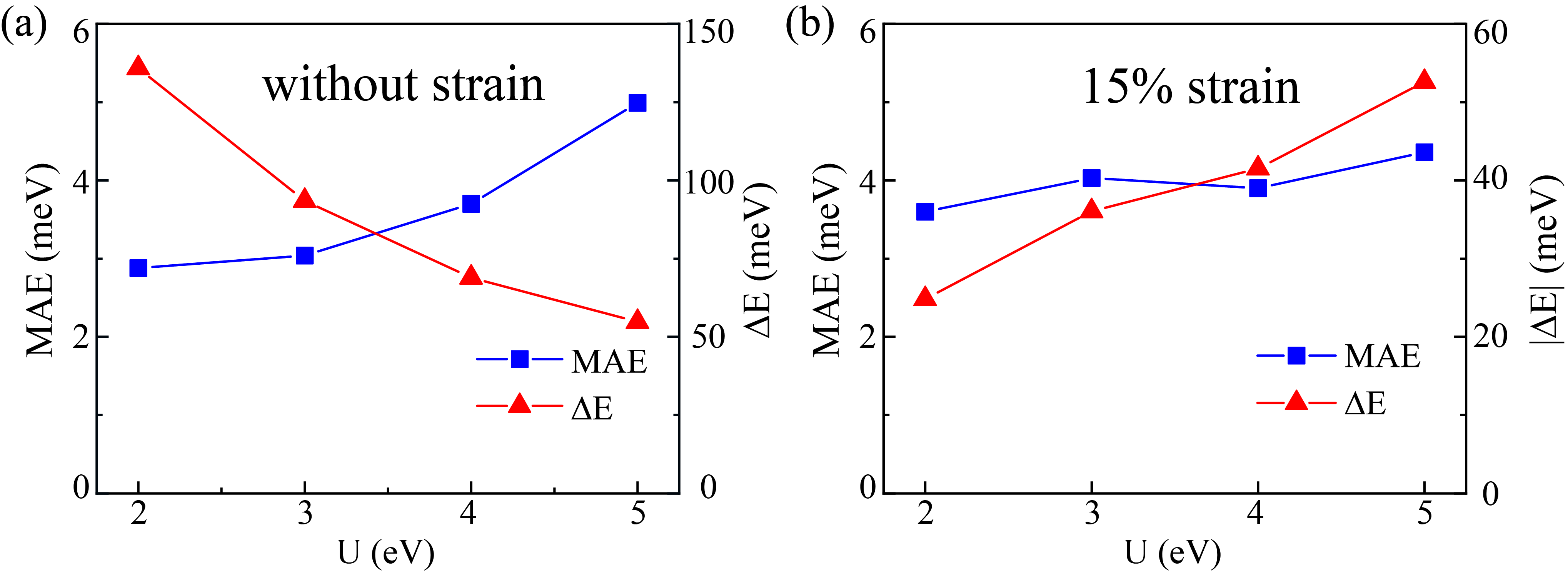}\\
  \caption{Electronic correlation $U$-dependent magnetic anisotropy energy (MAE) and energy difference ($\Delta$E) between antiferromagnetic and ferromagnetic configurations (a) without strain and (b) with 15$\%$ tensile strain. The results are calculated by GGA + SOC + $U$ method. }\label{fig8}
\end{figure*}

The electronic correlation effect is important for $3d$ orbitals in transition-metal compounds, so our above DFT calculations are carried out with $U$ = 4 eV. In order to examine the influence of different $U$ values on the magnetic ground states, we have studied the magnetic sate of Fe$_2$Cl$_3$I$_3$ for the cases without strain and with 15$\%$ tensile strain with parameter $U$ from 2 to 5 eV. As shown in Fig.\ref{fig8}, when there is no strain applied, with the increase of $U$ value, although MAE increases and $\Delta$E decreases, the signs of both preserve, indicating the unchanged AFM ground state with out-of-plane magnetization of sublattice. Similarly, the FM ground state along z-axis magnetization with 15$\%$ strain is not changed when different $U$ values are employed. Thus, our results about the magnetic phase transition with respect to the strain are robust against $U$ values.

\section{Conclusion}

By first-principles calculations, we have proposed a new 2D magnetic Janus semiconductor--Fe$_2$Cl$_3$I$_3$. which was revealed to exhibit the zigzag AFM ground state with out-of-plane magnetic direction. In contrast to non-Janus materials, the inversion symmetry breaking usually occurs in Janus materials. This can induce the intrinsic electric polarization and enhanced spin-orbital coupling. Fe$_2$Cl$_3$I$_3$ was found to possess a spontaneous polarization along the z-axis. Furthermore, we have also investigated the effect of biaxial strain on the ground state properties of Fe$_2$Cl$_3$I$_3$, and a magnetic phase transition including the antiferromagnetic-ferromagnetic transition and the change of magnetization direction was obtained. Both magnetic and electric polarization were observed in Fe$_2$Cl$_3$I$_3$ under the biaxial strain. A phase diagram based on the spin-spin interactions with the single-ion anisotropy term was proposed to interpret the magnetic phase transition. Our findings not only expose a new stable 2D magnetic Janus semiconductor, but also reveal the highly sensitive strain-controlled magnetic states, and thus highlight the 2D magnetic Janus semiconductor as a new platform to design spintronic materials.

%\begin{acknowledgement}
\section{Acknowledgements}
The authors thank Peng Fan for valuable discussions on the Monte Carlo simulation. This work is supported in part by the National Key R$\&$D Program of China (Grant No. 2018YFA0305800), the Strategic Priority Research Program of Chinese Academy of Sciences (Grant No. XDB28000000), the National Natural Science Foundation of China (Grant No. 11834014), and Beijing Municipal Science and Technology Commission (Grant No. Z191100007219013).
B.G. is also supported in part by the National Natural Science Foundation of China (Grant No. Y81Z01A1A9), the Chinese Academy of Sciences (Grant No. Y929013EA2), the University of Chinese Academy of Sciences (Grant No.110200M208), the Strategic Priority Research Program of Chinese Academy of Sciences (Grant No.XDB33000000), and the Beijing Natural Science Foundation(Grant No. Z190011).

Zhen Zhang and Jing-Yang You contributed equally to this work.
%\end{acknowledgement}

%\bibliographystyle{apsrev4-2}
%\bibliography{ref}


\end{document}